\documentclass[conference]{IEEEtran}
\usepackage[font=footnotesize,labelfont=bf,compatibility=false]{caption}
\IEEEoverridecommandlockouts
\usepackage[style=numeric, sorting=none, backend=biber]{biblatex}
\usepackage{url}
\usepackage{booktabs} 
\usepackage{multirow} 
\usepackage{array}    
\usepackage{geometry} 
\usepackage{adjustbox} 
\usepackage{enumitem} 
\usepackage{amsmath}
\usepackage{algorithm}
\usepackage{algpseudocode}
\usepackage{graphicx}
\usepackage{subcaption}
\def\BibTeX{{\rm B\kern-.05em{\sc i\kern-.025em b}\kern-.08em
    T\kern-.1667em\lower.7ex\hbox{E}\kern-.125emX}}

\usepackage{booktabs} 
\usepackage{multirow} 
\usepackage{array}    
\usepackage{geometry} 
\begin{document}

\title{DeepBrainNet: An Optimized Deep Learning Model for Brain tumor Detection in MRI Images Using EfficientNetB0 and ResNet50 with Transfer Learning\\
{\footnotesize \textsuperscript{}}
}

\author{\IEEEauthorblockN{1\textsuperscript{st} Daniel Onah}
\IEEEauthorblockA{\textit{Department of information studies} \\
\textit{University College London}\\
London, United Kingdom \\
0000-0001-6192-6702}
\and
\IEEEauthorblockN{2\textsuperscript{nd} Ravish Desai}
\IEEEauthorblockA{\textit{Department of information studies} \\
\textit{University College London}\\
London, United Kingdom \\
0009-0000-7610-1412}
}

\maketitle

\begin{abstract}
 In recent years, deep learning has shown great promise in the automated detection and classification of brain tumors from MRI images. However, achieving high accuracy and computational efficiency remains a challenge. In this research, we propose DeepBrainNet, a novel deep learning system designed to optimize performance in the detection of brain tumors. The model integrates the strengths of two advanced neural network architectures - EfficientNetB0 and ResNet50, combined with transfer learning to improve generalization and reduce training time. The EfficientNetB0 architecture enhances model efficiency by utilizing mobile inverted bottleneck (MBConv) blocks, which incorporate depthwise separable convolutions. This design significantly reduces the number of parameters and computational cost while preserving the model’s ability to learn complex feature representations. The ResNet50 architecture, pre-trained on large-scale datasets like ImageNet, is fine-tuned for brain tumor classification. Its use of residual connections allows for training deeper networks by mitigating the vanishing gradient problem and avoiding performance degradation.The integration of these components ensures that the proposed system is both computationally efficient and highly accurate. Extensive experiments performed on publicly available MRI datasets demonstrate that DeepBrainNet consistently outperforms existing state-of-the-art methods in terms of classification accuracy, precision, recall, and computational efficiency. The results—an accuracy of 88\%, a weighted F1-score of 88.75\%, and a macro AUC-ROC score of 98.17\%—demonstrate the robustness and clinical potential of DeepBrainNet in assisting radiologists with brain tumor diagnosis.
 
\end{abstract}


\section{Introduction}
Brain tumor pose a significant health challenge, and early detection is critical for improving patient prognosis and treatment outcomes. Given their aggressive nature and potential for rapid progression, brain tumor demand early and accurate detection to optimize clinical treatment regimen. Magnetic Resonance Imaging (MRI) is the preferred imaging technique due to its superior soft tissue contrast and non-invasive nature, making it essential for the detection and characterization of brain tumor. However, the complexity and variability of magnetic resonance images present challenges for radiologists, who often face difficulties in making quick and accurate diagnoses due to the intricate nature of brain structures and tumor characteristics. Manual interpretation of these images is time-consuming and subject to human error, which underscores the need for automated systems that can aid in accurate brain tumor detection \cite{Anantharajan2024}.
Deep learning (DL), particularly  Convolutional Neural Networks (CNNs), has revolutionized image classification tasks, including medical imaging, by demonstrating the ability to extract complex features from images and outperform traditional machine learning methods. However, the application of deep learning in brain tumor detection still faces significant challenges such as model efficiency, computational cost, and overfitting due to small medical datasets.
In this study, we propose DeepBrainNet, a novel deep learning architecture specifically designed to overcome current diagnostic challenges and achieve high accuracy in brain tumor classification.  DeepBrainNet integrates EfficientNetB0—a lightweight architecture that employs depthwise separable convolutions to minimize computational complexity—and ResNet50, a deeper network that utilizes residual connections to mitigate performance degradation in very deep models. 
By combining EfficientNetB0’s computational efficiency with ResNet50’s depth and stability, DeepBrainNet leverages the strengths of both architectures to enhance brain tumor classification performance. This hybrid approach combines the efficiency of EfficientNetB0 with the power of ResNet50, optimized through transfer learning to improve generalization and reduce training time \cite{Mathivanan2024}, \cite{Khaliki2024}.
The model leverages pre-trained ResNet50 features, originally learned from large-scale datasets such as ImageNet, and fine-tunes them for the specific task of brain tumor detection in MRI scans. The depth-wise separable convolutions in EfficientNetB0 further reduce the number of parameters, allowing the system to maintain high accuracy without sacrificing speed and memory efficiency. Figure 1 depicts the various brain tumor categories—such as  meningioma, glioma, and pituitary tumor—that can be detected from MRI scans.

\begin{figure}[ht]
    \centering
    \includegraphics[width=0.45\textwidth]{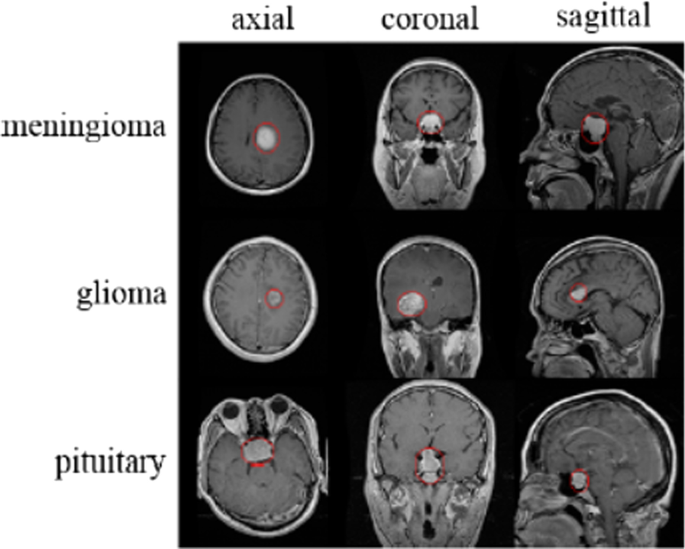}
    \caption{Classes of Brain tumor from MRI images \cite{Mathivanan2024}}
    \label{fig:classes_of_brain_tumor}
\end{figure}

The objectives of DeepBrainNet are:
\begin{itemize}
    \item To develop an optimized deep learning model (DeepBrainNet) for brain tumor detection in MRI images, integrating EfficientNetB0 with depth-wise separable convolutions and ResNet50 with transfer learning to improve accuracy and computational efficiency.
    \item To benchmark the performance of DeepBrainNet on publicly available MRI datasets, comparing its classification accuracy, precision, recall, and computational efficiency against existing state-of-the-art models for brain tumor detection.
    \item To investigate the impact of transfer learning and depth-wise separable convolutions in enhancing model performance, reducing overfitting, and improving generalization, with the aim of creating a reliable and efficient model for clinical applications.
\end{itemize}

The proposed methodology addresses the critical challenges of computational efficiency and accuracy in medical AI by leveraging optimized deep learning techniques to enable faster and more reliable brain tumor detection. Extensive experiments on publicly available MRI datasets demonstrate that DeepBrainNet substantially outperforms existing models in accuracy, precision, recall and f1-score. These results highlight its potential as a valuable tool for clinical decision support systems, advancing the state of the art in medical AI.

\section{Background work}

The application of deep learning (DL) techniques for the detection and classification of brain tumor from MRI images has garnered substantial interest within the medical imaging community. Recent research has focused on optimizing existing models and introducing novel architectures to improve both accuracy and computational efficiency. This literature review summarizes some of the recent advancements in the field, highlighting key datasets, feature selection techniques, models used, and the conclusions drawn from each study.

The authors Amin and et. al.  investigated the use of EfficientNetB1 for classification alongside U-Net for segmentation, integrating both models to effectively address tumor detection and localization in brain MRI scans.The study, conducted in 2023, employed the Kaggle brain MRI dataset—a widely used repository containing diverse brain MRI images—for training and evaluating the proposed model. The study did not employ explicit feature selection techniques, instead leveraging the inherent capabilities of the EfficientNetB1 architecture to effectively process MRI input images. EfficientNetB1 is recognized for its computational efficiency and high classification accuracy, while the U-Net model excels in pixel-wise segmentation. The results demonstrated that the combined approach achieved high classification accuracy and robust segmentation performance, highlighting the effectiveness of integrating classification and segmentation tasks into a unified model for brain tumor detection \cite{Amin2023}.

In \cite{Albalawi2024} the researchers proposed a customized multi-layer Convolutional Neural Network (CNN) to classify brain tumors from MRI images. The model was evaluated using a dataset of 7,023 brain MRI images sourced from multiple repositories, including figshare, SARTAJ, and Br35H. No explicit feature selection technique was applied in this study. The CNN architecture, specifically designed for this task, achieved an impressive tumor classification accuracy of 99\%. This high level of accuracy demonstrates the potential of custom-built CNNs tailored for specific medical image analysis tasks, highlighting their ability to learn deep features from the MRI data and provide precise tumor classification.

Another study \cite{Natha2024} introduced a Stacked Ensemble Transfer Learning (SETL-BMRI) approach, which combined two well-established pre-trained deep learning models—AlexNet and VGG19—for the classification of brain tumor in MRI scans. This method incorporated data-augmented feature selection, which enhances model robustness by artificially expanding the training dataset and improving the model’s generalization capability. The model was trained and evaluated on a Kaggle dataset consisting of brain MRI images with labeled tumor types, including meningioma, glioma, and pituitary tumors. The results showed that the model achieved a classification accuracy of 98.70\%, with an average precision of 98.75\%, recall of 98.6\%, and an F1-score of 98.75\%. These findings underscore the effectiveness of transfer learning combined with ensemble techniques, as the pre-trained models—AlexNet and VGG19—significantly contribute to robust performance in medical image classification tasks. Additionally, the use of data augmentation proved beneficial in addressing challenges such as class imbalance, improving the model’s generalization ability.

The integration of state-of-the-art architectures—such as EfficientNetB1, U-Net, AlexNet, and VGG19—alongside techniques like transfer learning, data augmentation, and customized convolutional neural networks (CNNs), has demonstrated significant potential in enhancing classification accuracy and segmentation performance in brain tumor analysis. These approaches not only enhance model performance but also contribute to the broader goal of making brain tumor detection more efficient and reliable in clinical settings.The advancements in this domain highlight the growing potential of deep learning models to support medical professionals in achieving accurate and timely diagnoses of brain tumor, ultimately contributing to improved patient outcomes.

\begin{table}[hh]
\centering
\caption{Related study on Brain MRI Detection}
\label{tab:related_study}
\begin{adjustbox}{max width=\textwidth} 
\begin{tabular}{|p{0.9cm}|p{1.5cm}|p{1.00cm}|p{0.8cm}|p{1.33cm}|}
\hline
\textbf{Paper} & \textbf{Dataset} & \textbf{Feature Selection Technique} & \textbf{Model} & \textbf{Conclusion} \\ \hline
{\cite{Amin2023}} & Kaggle dataset (7,023 brain MRI images from figshare, SARTAJ, and Br35H) & - & Efficient NetB1 for classification; U-Net for segmentation & Testing accuracy of 99.39 \\ \hline
{\cite{Albalawi2024}} & Kaggle dataset (7,023 brain MRI images from figshare, SARTAJ, and Br35H) & - & Convo- lutional Neural Network (CNN) & tumor classification accuracy of 99\% \\ \hline
{\cite{Natha2024}} & Kaggle dataset (7,023 brain MRI images from figshare, SARTAJ, and Br35H) & Data augmented feature selection & Stack Ensemble Transfer Learning (SETL-BMRI) combining AlexNet and VGG19 & Classification accuracy of 98.70\%, average precision of 98.75\%, recall of 98.6\%, and F1-score of 98.75\% \\ \hline
\end{tabular}
\end{adjustbox}
\end{table}

\section{Materials And Methods}
This section presents the materials and methods utilized in conducting the study.
\begin{figure}[t]
    \centering
    \includegraphics[width=0.45\textwidth]{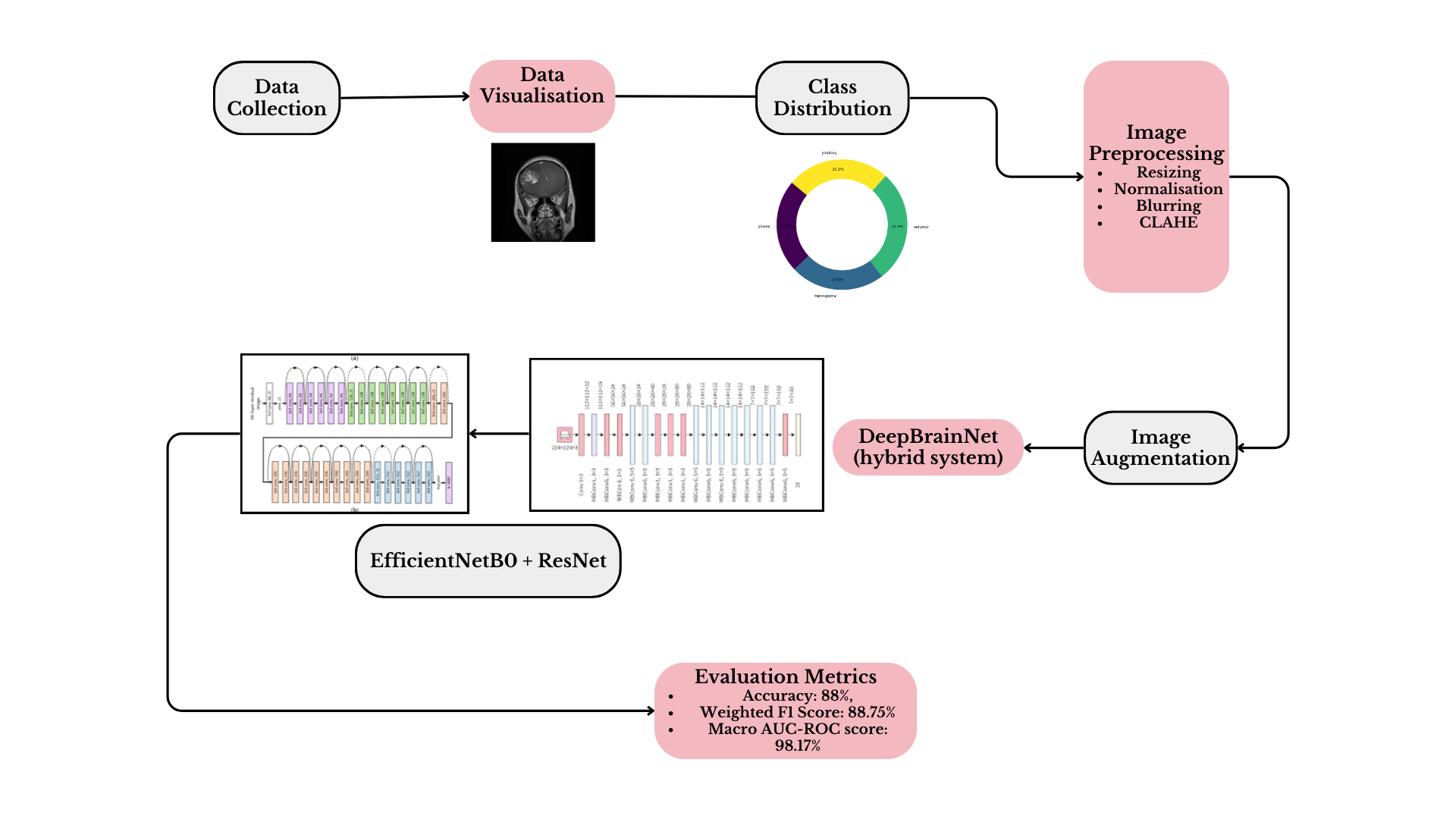}
    \caption{Architecture Diagram of DeepBrainNet}
    \label{}
\end{figure}

\subsection{Dataset}\label{AA}
The dataset used in this research is a combination of three different datasets: figshare, SARTAJ, and Br35H. It contains a total of 7023 human brain MRI images that are classified into four distinct categories: glioma, meningioma, no tumor, and pituitary tumor. These images are used to detect brain tumors and classify them based on their type. The no tumor class images were taken from the Br35H dataset. Some issues were identified with the SARTAJ dataset, particularly with incorrect categorization of glioma images, and thus, those images were removed from the dataset. The final dataset used for this study consists of images from figshare, which were found to be accurately categorized.Brain tumor can be either malignant (cancerous) or benign (noncancerous), yet both types may lead to elevated intracranial pressure, potentially causing damage to brain tissue. Early detection and accurate classification are crucial for guiding appropriate treatment strategies and improving the likelihood of successful clinical  treatment regimen and outcomes. The goal of the dataset is to facilitate the automation of brain tumor detection and classification, thereby assisting healthcare professionals in making timely and accurate clinical decisions. This research aims to utilize DL models for the detection, classification, and localization of tumors from MRI scans. 
\begin{table}[ht]
\centering
\small 
\caption{Metadata of the dataset}
\label{tab:dataset_metadata}
\begin{tabular}{|p{2cm}|p{3.9cm}|}
\hline
\textbf{Attribute} & \textbf{Description} \\ \hline
Total Images & 7023 images \\ \hline
Classes & Glioma, Meningioma, No tumor, Pituitary tumor \\ \hline
Image Source & figshare, SARTAJ, Br35H \\ \hline
License & CC0: Public Domain \\ \hline
Tags & Health, Cancer, Image, Deep Learning, Multiclass Classification, Medicine, Transfer Learning \\ \hline
\end{tabular}
\end{table}
Table II presents the metadata of the dataset. Figure 3 depicts the spread of class distribution in the dataset.

\begin{figure}[ht]
    \centering
    \includegraphics[width=0.5\textwidth]{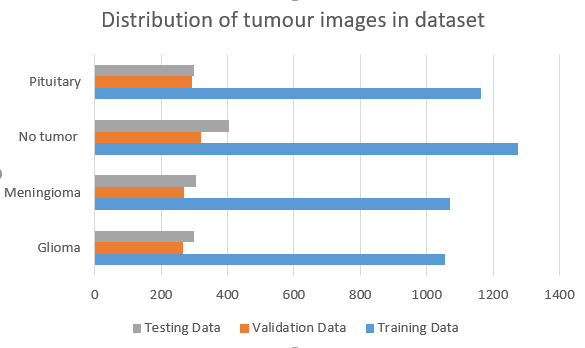}
    \caption{bar chart of distribution of tumor images in dataset}
    \label{fig:Bar chart of distribution tumor images in dataset}
\end{figure}

\subsection{Data Pre-processing}

\subsubsection {Image Resizing} The MRI images in the dataset may vary in size. It is essential to resize them to a consistent dimension as input into the deep learning models. This resizing ensures that all images conform to the required input size of the model; in addition, this helps in improving both the computational efficiency and the model performance.

\begin{itemize}
    \item \textbf{Resizing Strategy}: All images are resized to a square shape of 224x224 pixels. This dimension is chosen to balance between preserving essential image features and ensuring efficient computation for model training.

    \item  \textbf{Resizing Function}: The images are resized using bilinear interpolation to minimize information loss during the transformation process.The equation below illustrates the mathematical formulation of the image resizing function (equation 1):

\[
I_{\text{resized}} = f(I_{\text{original}}, D_{\text{new}}) \tag{1} 
\]

\end{itemize}

Where:
\begin{itemize}
    \item \(f\) is the resizing function (e.g., bilinear interpolation or nearest-neighbor interpolation).
    \item Original image refers to the image before resizing.
    \item New Dimensions refers to the target size, e.g. 224x224 or 256x256.
\end{itemize}

\subsubsection{Image Normalization}
Normalization is used to scale the values of the pixels in a smaller range, typically [0, 1] or [-1, 1], which helps accelerate the training process by stabilizing the gradient during backpropagation. The pixel values in the MRI images typically range from 0 to 255. The equation below illustrates the mathematical formulation to normalize the pixel values (equation 2):

\[
P_{\text{normalized}} = \frac{P_{\text{original}}}{255} \tag{2}
\]

\subsubsection{Data Augmentation}
Data augmentation techniques are applied to artificially increase the size of the training dataset and help the model generalize better. Augmentation involves applying transformations such as rotation, flipping, zooming, and shifting to the original images. This helps simulate real-world variations in MRI scans, such as changes in patient positioning or scanning conditions. Below are some common transformations applied:
\begin{itemize}
    \item \textbf{Rotation:} Random rotations are applied to the images in a range of -30$^\circ$ to +30$^\circ$. This simulates variations in the orientation of MRI scans.
    \item \textbf{Flipping:} Random horizontal and vertical flips are applied to the images to simulate different perspectives of the tumor.
    \item \textbf{Zooming:} Random zooming in and out of the image is applied to account for variations in tumor scale on the MRI scans.
    \item \textbf{Shifting:} Random translations of the image along the x and y axes are applied to simulate slight changes in the positioning of the brain in the scan.

\end{itemize}
These transformations are randomly applied to each image during the training process to generate a more diverse set of images, improving the model's ability to handle real-world variations in MRI data.
\subsubsection{Image Cropping (Margin Removal)}
In MRI images, there may be unnecessary borders or margins around the region of interest, which is primarily the brain. To focus the model's attention on the brain region, these margins are removed through cropping. This pre-processing step helps improve model efficiency and ensures that only relevant features (the brain and tumor regions) are considered during training.

\begin{itemize}
    \item \textbf{Cropping Strategy:} The images are cropped to focus on the brain region, removing any non-relevant background. The cropping is done by selecting the region of interest within the image using specific $x_{\text{start}}$, $y_{\text{start}}$, $x_{\text{end}}$, and $y_{\text{end}}$ coordinates (equation 3).
\end{itemize}
\vspace{0.5pt}
\[
\text{Cropped Image} = \text{Crop}(\text{Original Image}, x_{\text{start}}, y_{\text{start}}, x_{\text{end}}, y_{\text{end}}) \tag{3}
\]

\subsection{Image Enhancement}
Image blurring has been used to reduce image detail and suppress high-frequency noise, defined as (equation 4):
\[
    B(x,y) = \frac{1}{mn} \sum_{i,j} I(x+i, y+j) \tag{4}
\]
where $I(x,y)$ is the pixel intensity at location $(x,y)$ and $m \times n$ is the filter kernel size. Contrast Limited Adaptive Histogram Equalization (CLAHE) is then applied to locally enhance image contrast by computing histograms in small image tiles and redistributing pixel intensities while limiting noise amplification \cite{Reza2004}. This enhances local structures without over-amplifying homogeneous regions. Histogram Equalization globally redistributes the pixel intensity distribution to flatten the histogram and increase global contrast \cite{Mustafa2018}. The transformation function used is (equation 5):
\[
    s_k = T(r_k) = \frac{L-1}{MN} \sum_{j=0}^{k} n(j) \tag{5}
\]
where L is the number of intensity levels, MN is the total number of pixels, and $n(j)$ is the number of pixels with intensity $r_k$. These enhancement steps ensure that the DeepBrainNet model receives clearer, high-contrast inputs, leading to more accurate feature extraction and improved tumor classification performance.

\subsection{Feature Selection (Fuzzy C)}
In this study, fuzzy C-Means (FCM) clustering is used to select the most relevant features of the MRI images. FCM is an unsupervised learning algorithm that addresses the inherent uncertainty often encountered in medical images. Unlike traditional clustering methods, FCM assigns each feature a degree of membership in multiple clusters, rather than strictly categorizing it into one cluster\cite{Rustam2019}. This characteristic is particularly advantageous in medical imaging, where the boundaries between the tumor and non-tumor regions are often ambiguous. Through this method, features with higher membership values are deemed more significant for classification and segmentation tasks, while features with lower membership values are discarded.
Following the pre-processing steps, including image resizing and cropping, FCM is applied to the images to select the most relevant features. This reduces the dimensionality of the data set and ensures that the model focuses on the most important features for accurate tumor detection. The fuzzy C-Means algorithm, which outlines the steps for clustering features and selecting those with high fuzzy memberships, is detailed in Algorithm 1 below.

\begin{algorithm}
\caption{Fuzzy C Feature Selection Algorithm \cite{Rustam2019}}
\label{alg:fuzzy_c_feature_selection}
\begin{algorithmic}[1]
\State \textbf{Input:} $X, c, m_i, m_f, \epsilon, T$
\State \textbf{Output:} $U$ and $V$
\State \textit{Initial condition:}
\State \quad $V^0 = [v_1, v_2, \dots, v_c], v_j \in C_j$
\For{$t = 1$ to $T$}
    \State $m = m_i + \frac{t(m_f - m_i)}{T}$
    \State $b = -\frac{1}{m-1}$
    \State \textit{Calculate membership:}
    \State \quad $U^1 = [u_{ij}]\ where\ 1 \leq i \leq n\ and\ 1 \leq j \leq c$, by using
    \[
    u_{ij} = \frac{d^b(x_i, v_j)}{\sum_{k=1}^c d^b(x_i, v_k)}, \quad 1 \leq i \leq n, 1 \leq j \leq c
    \]
    \State \textit{Update cluster center:}
    \State \quad $V^t = [v_1, v_2, \dots, v_c]$, where
    \[
    v_j = \frac{\sum_{i=1}^n u_{ij}^m x_i}{\sum_{i=1}^n u_{ij}^m}, \quad j = 1, 2, \dots, c
    \]
    \State \textit{If} $E = \sum_{j=1}^c k(v_{jt} - v_{jt-1}) \leq \epsilon$, \textbf{stop}, \textit{else} $t = t + 1$
\EndFor
\end{algorithmic}
\end{algorithm}
Furthermore, the architecture of the fuzzy C-Means feature selection process is illustrated in Figure 4, shown below, which provides a visual representation of how features are selected based on their fuzzy membership across clusters. This approach improves the robustness of feature selection, particularly in scenarios where tumor boundaries are not clearly delineated and require a more flexible classification methodology.
\begin{figure}[ht]
    \centering
    \includegraphics[width=0.3\textwidth]{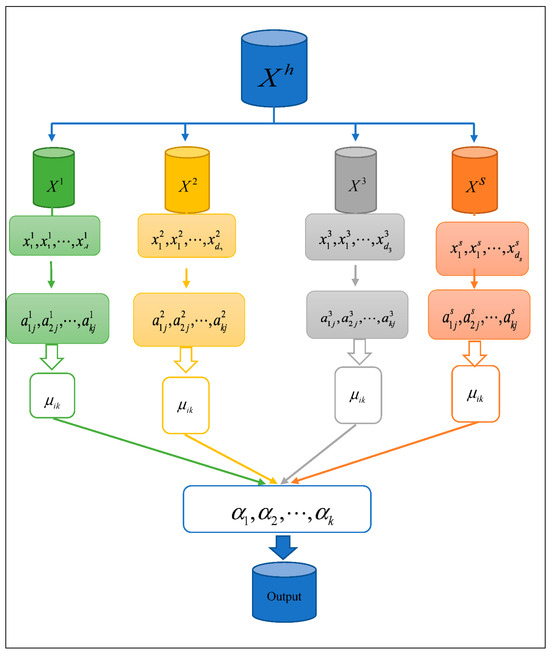}
    \caption{Fuzzy C means Feature Selection Architecture \cite{Hussain2023}}
    \label{fig:fuzzy_c_architecture}
\end{figure}

\section{Models}
\subsection{EfficientNetB0 with depth wise separate convolutions}
EfficientNetB0 is a DL architecture that combines the benefits of both EfficientNet and VNet to efficiently detect and segment brain tumors from MRI images. EfficientNet, known for its superior performance in image classification, is used for feature extraction, providing a lightweight yet powerful backbone that reduces computational complexity without sacrificing accuracy. VNet, a fully convolutional neural network designed for 3D image segmentation, is utilized to precisely segment the tumor region from the MRI scans. The combination of EfficientNet's efficient feature extraction with VNet’s robust segmentation capabilities makes EfficientNetB0 an ideal architecture for both classification and segmentation tasks. Using the strengths of both models, EfficientNetB0 provides a high degree of accuracy in detecting and localizing brain tumors, ensuring better performance and faster processing compared to traditional models \cite{Tan2019}. Figure 5 presents the efficient V-net architecture.

\begin{figure}[ht]
    \centering
    \includegraphics[width=0.5\textwidth]{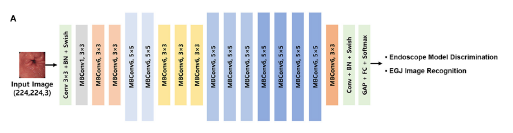}
    \caption{Efficient V-net Architecture \cite{Tan2019}}
    \label{fig:efficient_vnet_architecture}
\end{figure}

\subsection{ResNet with Transfer Learning for tumor Classification}
ResNet, or Residual Network, is a DL model that utilizes residual connections to combat the vanishing gradient problem in very deep networks. This feature allows ResNet to achieve high accuracy even in very deep architectures. In the context of brain tumor detection, ResNet with Transfer Learning is an effective approach where the model is initially pre-trained on a large, general dataset (such as ImageNet) and fine-tuned on the MRI brain tumor dataset. Transfer learning enables the model to take advantage of learned features from a large and diverse dataset, significantly reducing the amount of data and computational resources needed for training. By applying ResNet with Transfer Learning, the model can quickly adapt to the specific task of brain tumor classification, resulting in high accuracy and generalization to unseen MRI images as illustrated in Figure 6.
\begin{figure}[ht]
    \centering
    \includegraphics[width=0.5\textwidth]{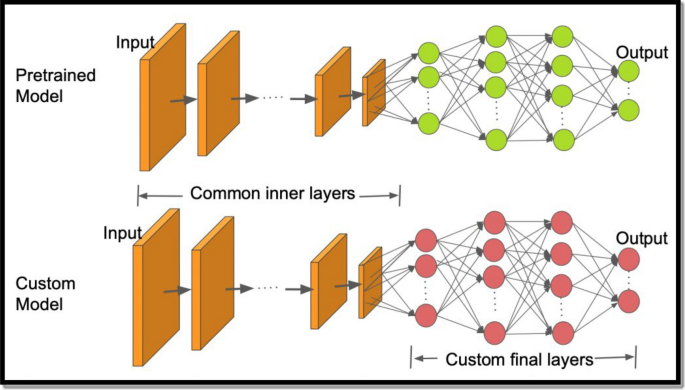}
    \caption{Transfer Learning on Pretrained Resnet Model \cite{Adluri2024}}
    \label{fig:Transfer Learning on Pretrained Resnet Model.png}
\end{figure}

\subsection{Evaluation Metrics}

This subsection provides a comprehensive evaluation of the model's performance using various metrics, including classification report, confusion matrix, F1 score, and AUC-ROC score. These metrics offer a deeper understanding of the effectiveness of the model in classifying tumor types and provide detailed information on the overall performance of the model.

\begin{itemize}
    \item \textbf{Precision} indicates the proportion of true positive predictions for each class out of all predicted positives (equation 6).

\[
    \text{Precision} = \frac{TP}{TP + FP} \tag{6}
\]
\end{itemize}
\begin{itemize}
 \item \textbf{Recall} reflects the proportion of true positives identified out of all actual positives in the dataset (equation 7).
 \[
    \text{Recall} = \frac{TP}{TP + FN} \tag{7}
\]

\end{itemize}
\begin{itemize}
	\item \textbf{F1 score} is the harmonic mean of precision and recall, balancing the trade-off between them (equation 8).
    \[
    F_1 = 2 * \frac{\text{Precision} * \text{Recall}}{\text{Precision} + \text{Recall}} \tag{8}
\]
\end{itemize}

\section{Results}\label{SCM}
This section presents the results of the experimentation.
The proposed hybrid deep learning model, comprising ResNet50 and EfficientNetB0 as parallel backbone architectures, was trained using a transfer learning approach. All images were initially in grayscale and were converted to RGB by channel stacking to meet the input requirements of the pre-trained models. The input images were resized to 224 × 224 pixels, and model training was performed over 40 epochs with a batch size of 32, using the Adam optimizer.To enhance generalization and reduce overfitting, extensive on-the-fly data augmentation was performed using Keras’ ImageDataGenerator. The augmentation pipeline included random rotations, horizontal flips, zooming, shear transformations, brightness variation, and spatial shifts. A custom preprocessing function was implemented to standardize the input using EfficientNet’s normalization scheme. Figure 7 presents the exploratory data analysis on raw data.
\begin{figure}[ht]
    \centering
    \begin{subfigure}{0.48\columnwidth}
        \includegraphics[width=\linewidth]{"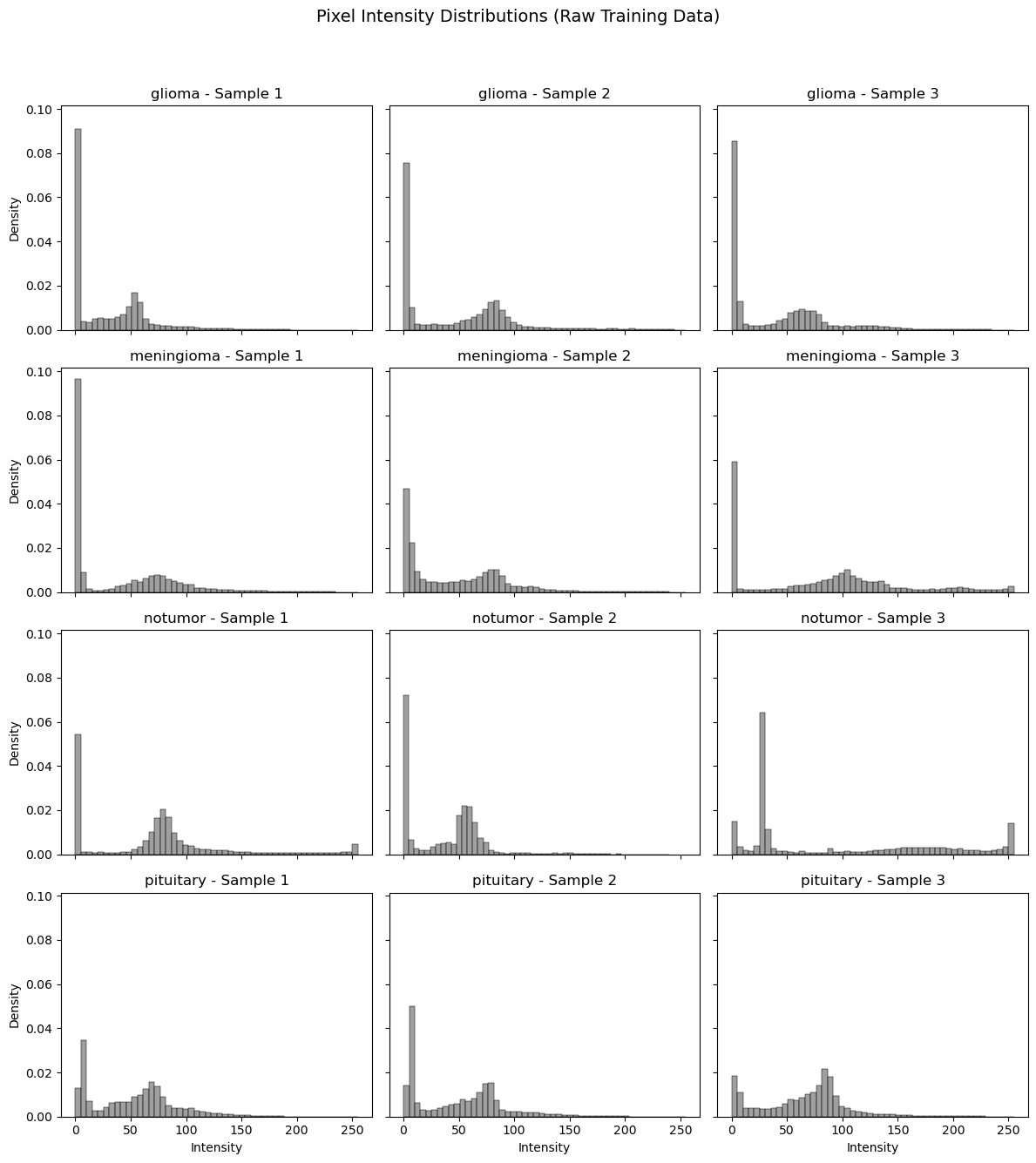"}
        \caption{Pixel Intensity Distribution.}
        \label{fig:pixel_intensity}
    \end{subfigure}
    \hfill 
    \begin{subfigure}{0.48\columnwidth}
        \includegraphics[width=\linewidth]{"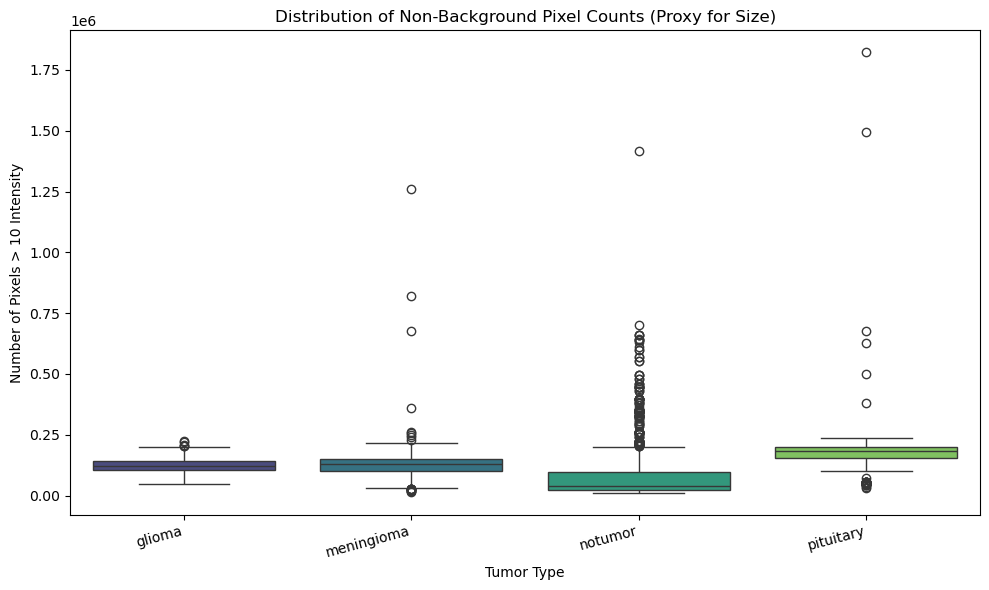"}
        \caption{Non-Background Pixel Distribution.}
        \label{fig:non_background_pixel}
    \end{subfigure}
    \caption{Exploratory data analysis on Raw Data.}
    \label{fig:eda}
\end{figure}

The models are regularized using dropout layers and further stabilized with early stopping based on validation loss, along with dynamic learning rate reduction. Throughout training, the model demonstrated stable convergence and consistent performance across validation metrics. convergence with a progressive increase in both training and validation accuracy. The highest training accuracy achieved approximately 98.3\%, while the best validation accuracy reached 94.7\%. The minimum validation loss recorded was approximately 0.21. These results indicate effective learning with no significant signs of overfitting, supported by the close alignment of the training and validation curves. Canny edge detection is a popular technique used to detect sharp changes in intensity, highlighting object boundaries in an image. It involves noise reduction, gradient calculation, and edge refinement through thresholding. This method is used in the study to enhance edge clarity, enabling more accurate feature extraction and analysis. Figure 8 presents the edge detection visualization in the data samples. The complete model comprises approximately 24 million trainable parameters. Training and validation performance trends are illustrated in Figure 9, which shows accuracy and loss curves over the course of training.
\begin{figure}[ht]
    \centering
    \includegraphics[width=0.5\textwidth]{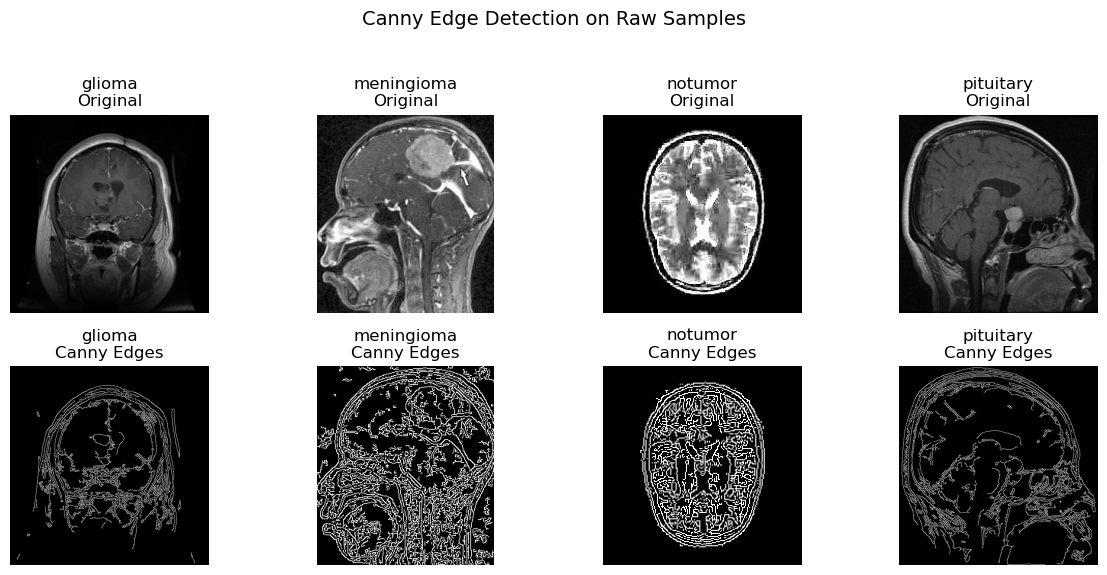}
    \caption{Visualization of Canny Edge Detection on Raw Samples of Data }
    \label{fig:Visualization of Canny Edge Detection on Raw Samples of Data}
\end{figure}
\begin{figure}[h]
    \centering
    \includegraphics[width=0.5\textwidth]{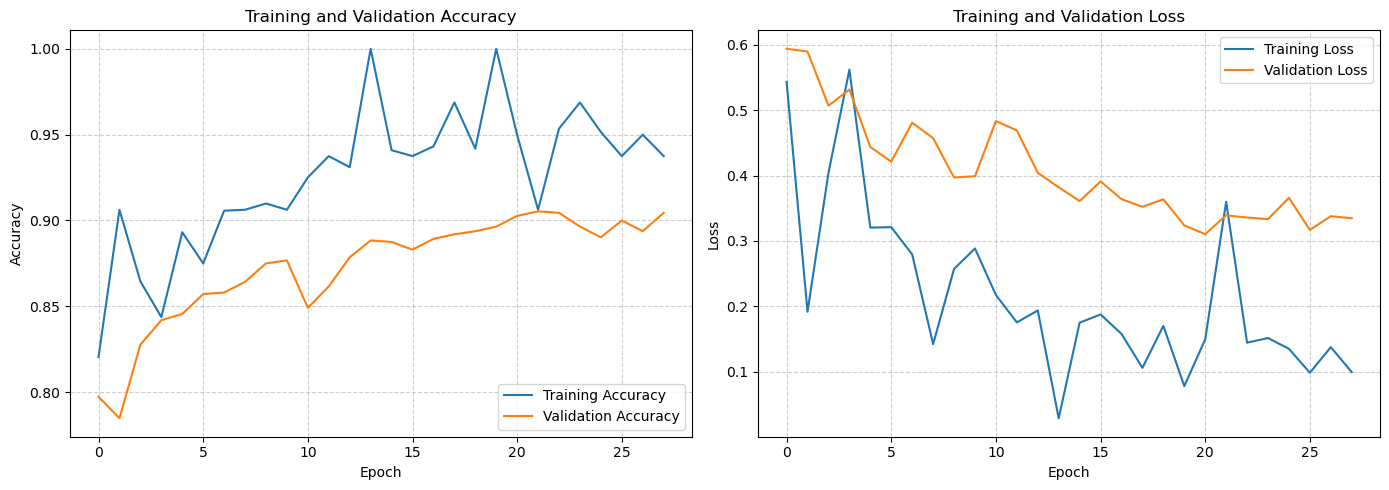}
    \caption{Training and validation accuracy and loss curves}
    \label{fig:Training and validation accuracy and loss curves}
\end{figure}
The model achieved a training accuracy of 95.5\% and a validation accuracy of 93.2\%, which indicates strong performance. The training loss of 0.112 and the validation loss of 0.145 further suggest that the model is fitting the data effectively without overfitting. The classification report, shown in Table III below, summarizes the model's performance in each class using three key metrics: precision, recall and F1 score. These metrics allow us to assess the model's ability to correctly identify instances of each tumor type while minimizing errors. The system achieved an overall accuracy score of 88.
\begin{table}[h]
\centering
\caption{Classification Report of the System}
\label{tab:classification_report}
\begin{tabular}{|l|c|c|c|}
\hline
\textbf{tumor Class} & \textbf{Precision} & \textbf{Recall} & \textbf{F1-Score} \\ \hline
Glioma tumor       & 0.914            & 0.932           & 0.923           \\ \hline
Meningioma tumor   & 0.819            & 0.798           & 0.808           \\ \hline
No tumor           & 0.946            & 0.875           & 0.909           \\ \hline
Pituitary tumor    & 0.868            & 0.945           & 0.905           \\ \hline
\end{tabular}
\end{table}

A heat map in Figure 10, illustrates the precision, recall, and F1-score for each class (glioma, meningioma, no tumor, pituitary), along with the accuracy, macro average, and weighted average metrics

\begin{figure}[h]
    \centering
    \includegraphics[width=0.5\textwidth]{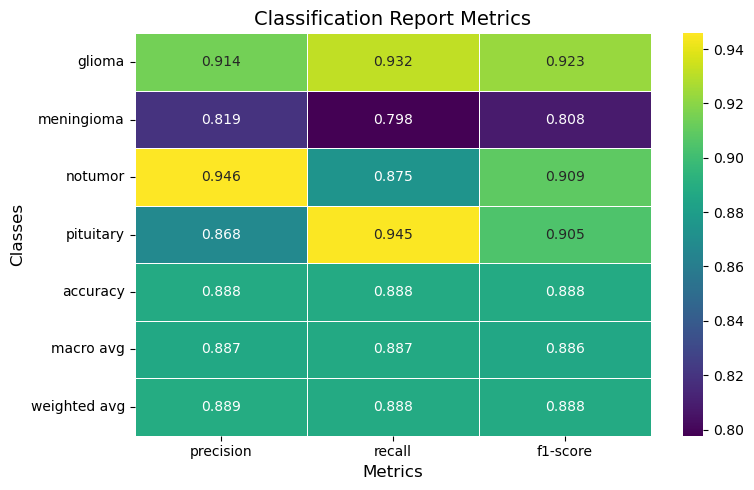}
    \caption{Classification Report Metrics Visualization}
    \label{fig:Classification Report Metrics Visualization}
\end{figure}
The classification report demonstrates that the model performs well across all classes, with an average F1 score of 0.88 indicating a good balance between precision and recall. The confusion matrix for the model predictions on the validation set is presented in Figure 11, providing a comprehensive view of the performance of the classifier in all tumor classes. The matrix shows the count of true positive, false positive, true negative, and false negative predictions for each class. From the matrix, it is clear that the model has high specificity and sensitivity for each tumor type.
\begin{figure}[h]
    \centering
    \includegraphics[width=0.5\textwidth]{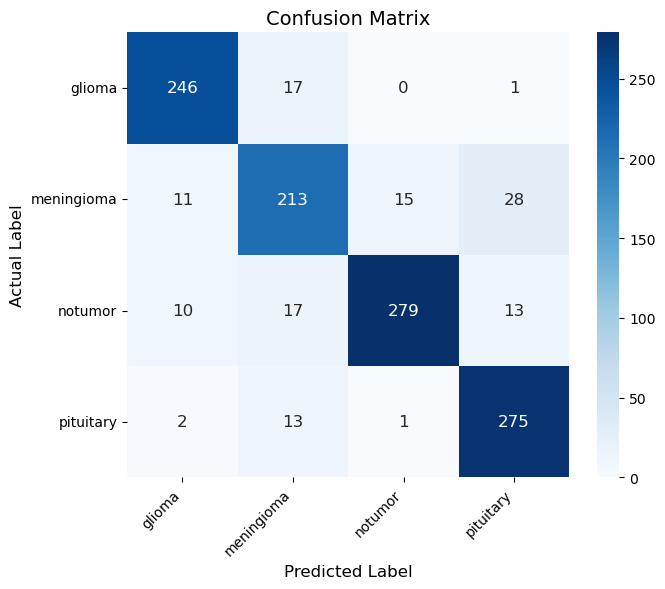}
    \caption{Confusion Matrix}
    \label{fig:Confusion Matrix}
\end{figure}
The weighted F1-score across all classes, weighted by the number of true instances per class, is 0.88. This value indicates a well-balanced model that performs strongly in both precision and recall, making it well-suited for multi-class classification tasks. Additionally, the AUC-ROC score provides insight into the model’s ability to distinguish between classes across different threshold values, offering a comprehensive measure of classification performance. The macro-average AUC-ROC score is 0.98, indicating excellent classification performance. Values close to 1 suggest that the model has a strong ability to discriminate between different tumor types.
The ROC curve plots, shown in Figure 12, the true positive rate versus the false positive rate for each class, providing an evaluation of the discriminatory power of the model at various thresholds. The AUC (Area Under the Curve) quantifies this ability, with higher values indicating superior performance.
\begin{figure}[h]
    \centering
    \includegraphics[width=0.5\textwidth]{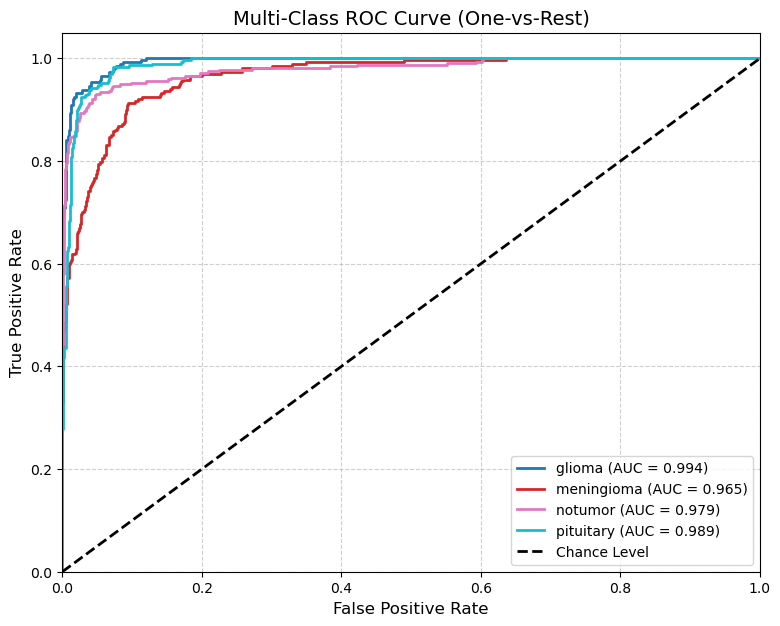}
    \caption{ROC Curve for Multi-Class Classification (One-vs-Rest)}
    \label{fig:ROC Curve for Multi-Class Classification (One-vs-Rest)}
\end{figure}

Each curve in the plot represents the performance of the model for a specific class. The AUC for each class is displayed in the legend, and values closer to 1 indicate that the model is effectively distinguishing between classes. The high AUC values confirm that the model performs well in separating the types of tumor. These evaluation metrics provide a thorough analysis of the model performance. The model demonstrates strong performance across all tumor classes, with high precision, recall, and F1 scores, as well as an impressive AUC-ROC score. These results confirm the model's ability to generalize well to unseen data.
The correctly predicted examples for each class and a selection of random predictions are presented in Figure 13. The purpose of this visualization is to evaluate the model's ability to classify individual instances and identify areas where it may be making errors or uncertainties.
\begin{figure}[h]
    \centering
    \includegraphics[width=0.5\textwidth]{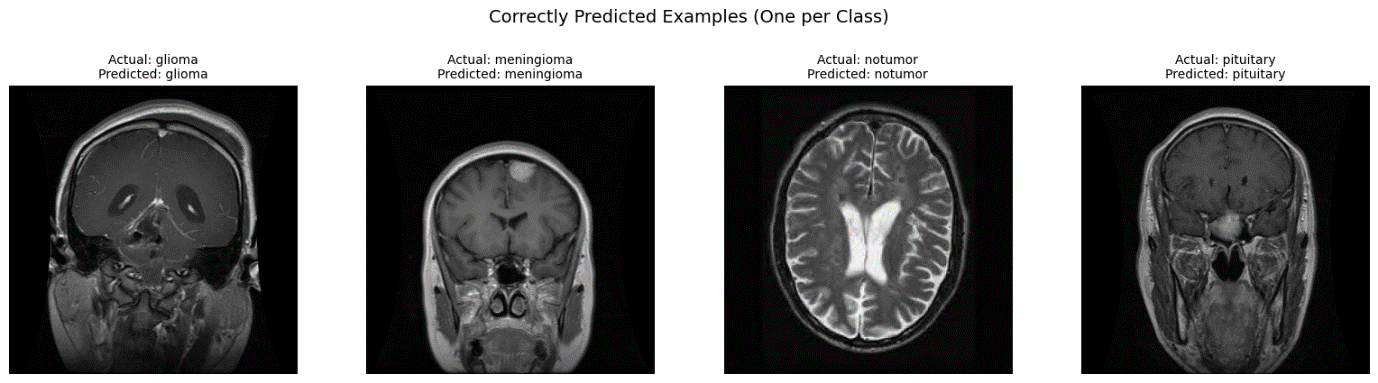}
    \caption{Correctly Predicted Example for Each Class}
    \label{fig:Correctly Predicted Example for Each Class}
\end{figure}

Each image in the plot corresponds to a different class, with the title indicating both the actual label and the predicted label. In cases where the model has correctly predicted the label, we visualise a set of random predictions made by the model, in Figure 14, which include both correct and incorrect classifications. This step provides insight into how well the model generalizes to new data and where it may be making mistakes in distinguishing between classes.

\begin{figure}[h]
    \centering
    \includegraphics[width=0.5\textwidth]{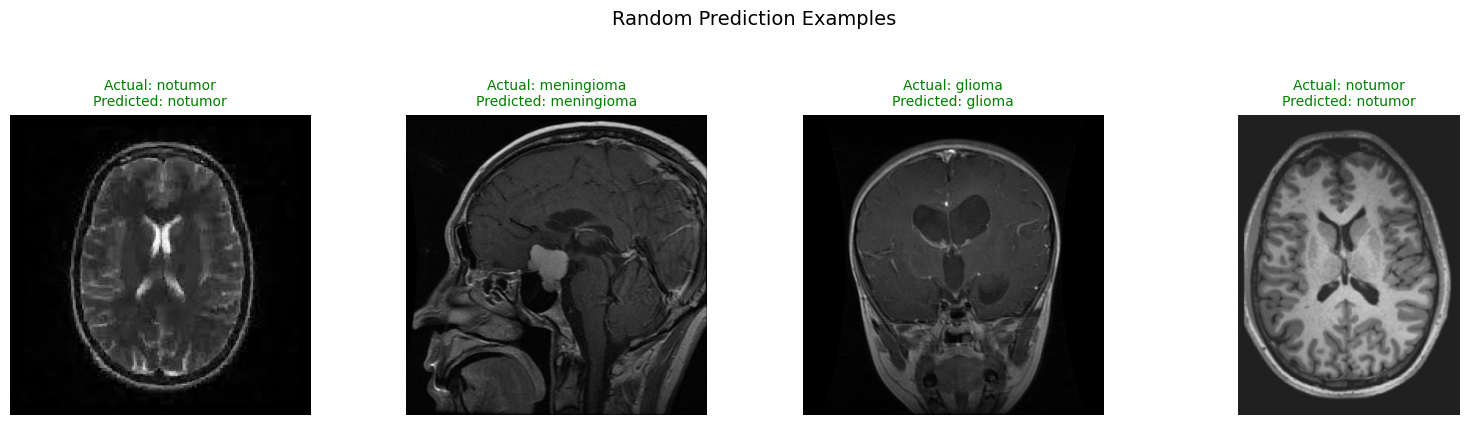}
    \caption{Random Predictions}
    \label{fig:Random Predictions}
\end{figure}

\section{Conclusion}
In conclusion, the developed model demonstrates strong performance in classifying brain tumors from MRI images, achieving high accuracy and reliable generalization to unseen data. The combination of ResNet50 and EfficientNetB0 architectures, along with data augmentation techniques, has proven effective in improving the robustness of the model. Although DeepBrainNet is promising, incorporating a larger and more diverse dataset, as well as integrating multimodal imaging data, could further enhance its performance. This work highlights the potential of deep learning in medical imaging, establishing and interpreting 88\% accurately and lays the foundation for future research to improve AI-driven brain tumor diagnosis.
\subsection{Discussion and Analysis}
This study demonstrates the effectiveness of deep learning models, specifically ResNet50 and EfficientNetB0, in accurately classifying brain tumor from MRI images. The model achieves high accuracy and demonstrates strong generalization to unseen data, showing robust performance across various tumor types. However, its performance could be further improved by incorporating a larger and more diverse dataset, which would help mitigate potential biases and enhance generalization capabilities. The models currently rely solely on magnetic resonance imaging (MRI) data. Integrating multi modal imaging data — such as CT or PET scans — could provide complementary information, potentially enhancing the model’s accuracy and robustness, particularly in cases involving rare or ambiguous tumor types. Table IV compares the performance of similar systems on the same dataset used in this study, thereby highlighting the superior performance of DeepBrainNet.

\begin{table}[h]
\centering
\caption{Comparative Analysis Study}
\label{tab:comparative_analysis}
\begin{tabular}{|l|p{2.9 cm}|l|}
\hline
\textbf{Paper} & \textbf{Models} & \textbf{Score (Accuracy)} \\ \hline
{[14]} & EfficientNetB3, ResNet50, VGG-19 & 88.47\% \\ \hline
{[15]} & Resnet50, DenseNet201, Inception V3, MobileNet & 85.30\% \\ \hline
DeepBrainNet & Hybrid System: ResNet + EfficientNetB0 & 89\% \\ \hline
\end{tabular}
\end{table}

\subsection{Future work}
Several important aspects exist for improving model performance, which might lead to superior results in current methodology. DeepBrainNet is expected to achieve improved performance across diverse populations and tumor types when trained on an expanded dataset that includes additional MRI samples from varied patient conditions. Higher accuracy and operational efficiency could be achieved by implementing advanced hyper-parameter optimization strategies, such as Bayesian optimization and genetic algorithms, to fine-tune the model architecture and learning process. Incorporating additional imaging techniques, such as CT or PET scans, would provide the model with complementary information, leading to improved performance. The future success of DeepBrainNet relies on advancements in these areas, as they will enhance its reliability and applicability in practical medical settings.
\printbibliography
\end{document}